\documentstyle[emulateapj,epsfig,psfig]{article}

\begin{document}

\title{75 kpc trails of ionized gas behind two Irr galaxies in A1367
}
\author{G. Gavazzi$^1$, A. Boselli $^2$, L. Mayer$^3$,
J. Iglesias-Paramo$^2$, J. M.  V\' \i lchez$^4$ \& L. Carrasco$^{5,6}$}

\affil{
$^1$ Universit\`a degli Studi di Milano - Bicocca, P.zza delle scienze 3,
20126 Milano, Italy\\ 
$^2$ Laboratoire d'Astrophysique de Marseille, Traverse du Siphon, F-13376, Marseille
Cedex 12, France\\
$^3$ Department of Astronomy, University of Washington, Seattle\\
$^4$ Instituto de Astrof\' \i sica de Andaluc\' \i a, CSIC, Apdo. 3004, 18080,
Granada,  Spain\\
$^5$Instituto Nacional de Astrof\' \i sica, Optica y Electr\'onica,
Apartado Postal 51. C.P. 72000 Puebla, Pue., M\'exico\\
$^6$Observatorio Astron\'omico Nacional, UNAM, Apartado Postal 877, C.P. 22860, Ensenada B.C., M\'exico}
\begin{abstract}
In a 6h $\rm H\alpha$ exposure 
of the N-W region of the cluster of galaxies 
A1367 we discovered a 75 kpc cometary emission of ionized gas trailing behind two Irregular galaxies.
The $\rm H_{\alpha}$ trails correspond in position and length with tails of syncrotron
radiation. At the galaxy side opposite to the tails the two galaxies show bright HII regions
aligned along arcs, where the star formation takes place at the prodigeous rate of $\rm \sim 1 M\odot yr^{-1}$.
From the morphology of the galaxies and of the trailing material, we infer that the two galaxies
are suffering from ram pressure due to their high velocity motion through the cluster IGM.
We estimate that $\sim 10^9$ $\rm M\odot$ of gas, probably ionized 
in the giant HII regions, is swept out forming the tails.
The tails cross each other at some 100 kpc from the present galaxy location, indicating that
a major tidal event occurred some $\sim 5 \times 10^7$ yr ago.
We exclude that mutual harassment produced the observed morphology and
we show with numerical simulations that it could have marginally aided 
ram pressure stripping by loosening the potential well of the galaxies.
\end{abstract}


\section {Introduction}

\begin{figure}[!h]
\epsscale{0.7}
\plottwo{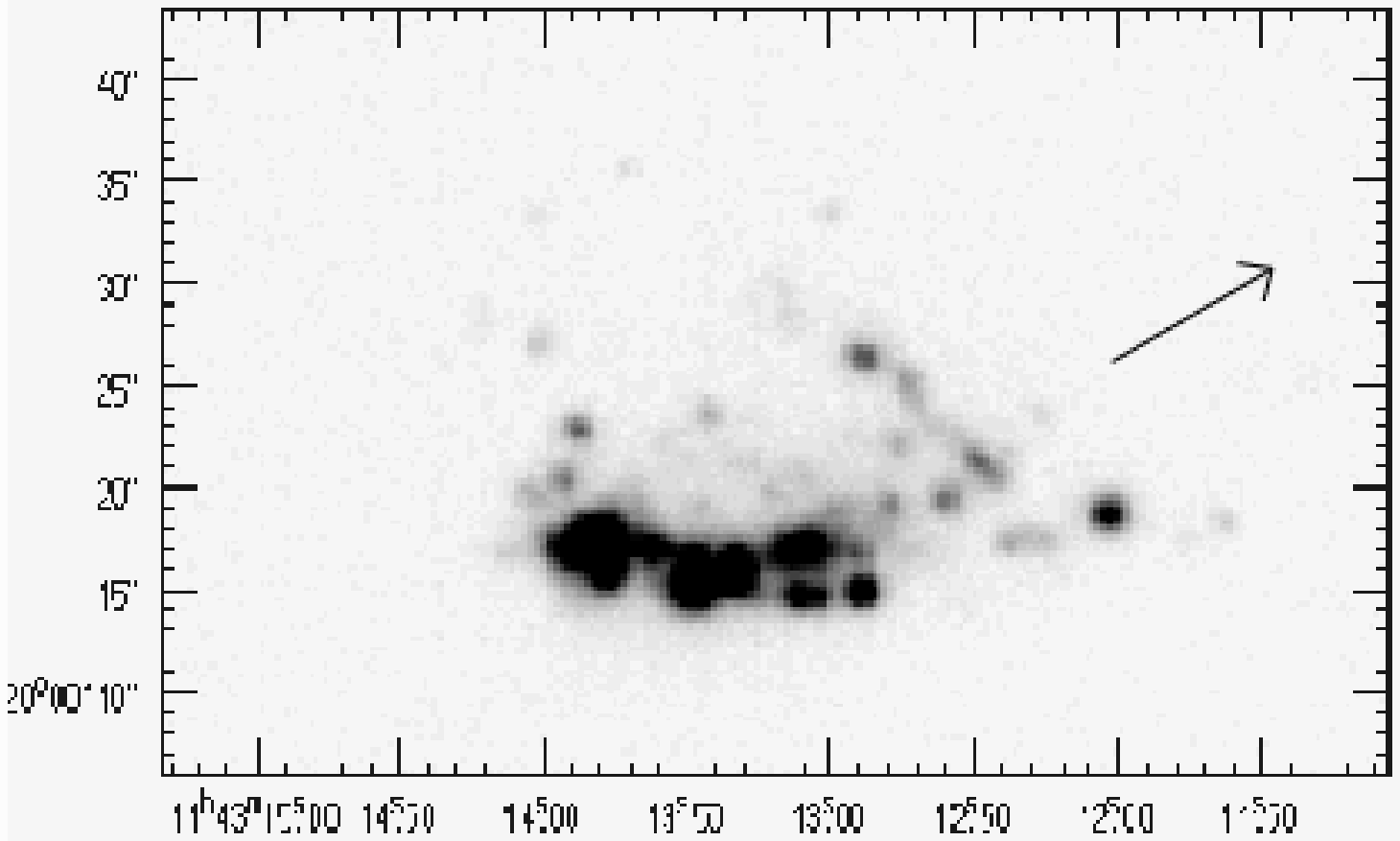}{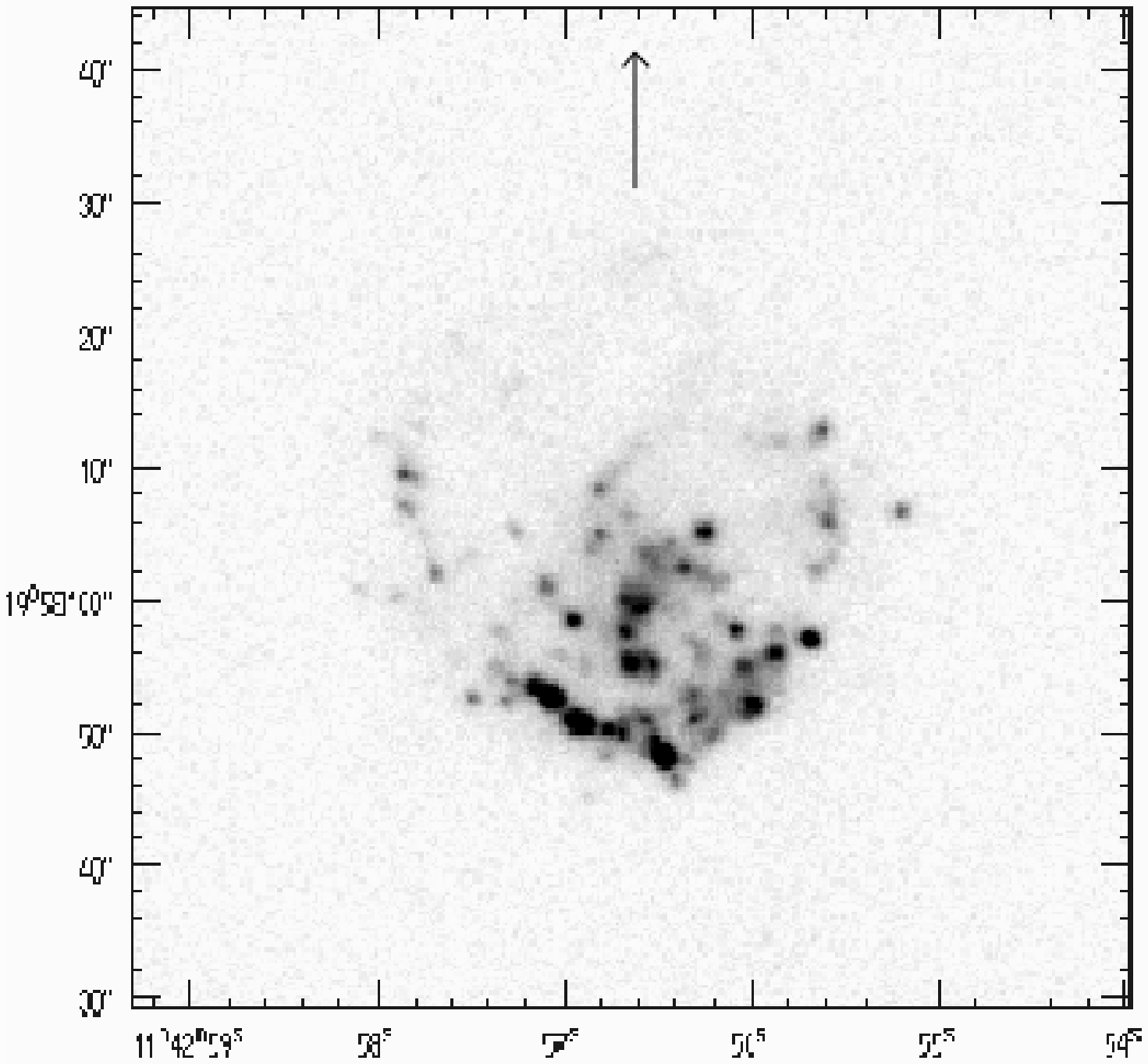}
\small{\caption{The $\rm H_{\alpha}+[NII]$ NET image taken with the WHT
with 0.85 arcsec seeing showing the bright parts of 97-079 (left) and 97-073 (right). (J2000)
celestial coordinates are given. The center of the cluster
is at S-E. An arrow marks the direction of the low brightness tail (see Fig. 2).}\label{97-073}}
\end{figure}

CGCG 97-073 and 97-079 (Zwicky et al, 1961-68) are 
Irregular members to the cluster of galaxies A1367. 
They have among the highest star formation rate found in Irregular galaxies:
$\rm \sim 1 M\odot yr^{-1}$ as derived from their
$\rm H_{\alpha}$ luminosity $\rm \sim 1.5\times10^{41} erg~sec^{-1}$ 
(Kennicutt, Bothun \& Schommer, 1984; Gavazzi et al. 1998). 
The star formation takes place in bright HII regions distributed
along curved paths on the galaxy periphery facing the cluster center (see Fig. \ref{97-073}). 
Both galaxies are significantly anemic on the opposite side. 
Radio continuum observations with the VLA
(Gavazzi \& Jaffe 1985, Gavazzi \& Jaffe 1987, Gavazzi et al. 1995),
revealed the "head-tail" appearence of the radio source associated with 
these two galaxies.
The tails extend up to 75 kpc (assuming $\rm H_0=75~km~s^{-1}~Mpc^{-1}$)
on the side opposite to the bright HII regions.
Observations in the 21 cm line of HI (Gavazzi 1989; Dickey \& Gavazzi 1991) revealed
that both galaxies have a slightly deficient
HI content, displaced in the direction marked by the radio continuum tails, 
as opposed to their H$_2$ content which appears normal in all respects
(Boselli et al. 1994). 
These asymmetries suggest that the two galaxies are experiencing ram pressure 
due to their high velocity motion through the IGM.
Abadi et al. (1999), Murakami \& Babul (1999) and Quilis et al. (2000) performed 
high-resolution hydrodynamical simulations
of galaxies subject to ram pressure stripping in rich clusters. In less than 1 Gyr their
galaxy models lose
all the gas as a result of ram pressure and viscous stripping when the density of the
IGM and the transit velocity are as high as in the Coma cluster. Extended gaseous tails form and
the gas is shocked at the leading edge of the galaxies,   considerably enhancing its density,  
thus most likely leading to intense star formation, as observed in 97-073
or in NGC 4522 (Kenney \& Koopmann 1999) in the Virgo cluster. \\
These phenomena might contribute significantly to 
the enrichment of the intergalactic
medium in clusters, which remains an unsettled issue (e.g. Madau, Ferrara \& Rees 2001; 
Mori, Ferrara \& Madau, 2001; Silich, et al. 2001; Recchi et al. 2001).

While taking deep (one hour exposure) $\rm H_{\alpha}$ images of the cluster A1367 with
the Wide Field Camera at the Isaac Newton Telescope (La Palma), we serendipitously discovered
a low surface brightness $\rm H_{\alpha}$ emission trailing behind 97-079, spatially 
coincident with the radio continuum tail. This emission was too faint to be measured and
required confirmation. We thus took more observations\footnote{Based on observations 
taken with the INT, WHT, NOT and SPM telescopes.
The INT, WHT and NOT are operated by the I.N.G. at the Spanish
Observatorio del Roque de Los Muchachos of the I.A.C. The SPM telescope
belongs to O.A.N de Mexico.} using narrow-band $\rm H_{\alpha}$
filters: 1h with the NOT and 4h with the SPM telescopes. Each individual observation
confirmed the existence of the trailing $\rm H_{\alpha}$.
The resulting stacked 6h exposure image, which we present in this Letter,
is sufficiently deep to allow a robust determination of the flux in the tail
of 97-079 and revealed an even fainter trail behind 97-073.
The cometary $\rm H_{\alpha}$ trails discovered in A1367 are the most extended low-brightness   
$\rm H_{\alpha}$ emission features ever detected. Other prominent 
examples are the the Magellanic stream (Weiner \& Williams, 1996) and the cloud in Leo
(Reynolds et al. 1986). 
\begin{table*}
\caption{The imaging instrumental set-up}
\label{Tab3}
\[
\begin{array}{p{0.15\linewidth}ccccccc}
\hline
\noalign{\smallskip}
{Telescope} & {Date} & {CCD} & {Pix}  & Filter  & Tint & seeing\\
            &        &       & arcsec & \AA     & sec  & arcsec\\
\noalign{\smallskip}
\hline
\noalign{\smallskip}
INT  & 26~Apr~2000 & 4 \times 2048 \times4100~EEV  & 0.33 & 6725~(80)   & 3600  & 1.7\\
INT  & 26~Apr~2000 & 4 \times 2048 \times4100~EEV  & 0.33 & (Gunn)~r'   & 900   & 1.7\\
SPM  & 23~Apr~2001 &  1024 \times1024~Thompson     & 0.38 & 6723~(80)   & 14400 & 1.9\\
SPM  & 23~Apr~2001 &  1024 \times1024~Thompson     & 0.38 & (Johnson)~R & 1800  & 1.9\\
NOT  & 25~Apr~2001 &  2048 \times2048~Loral        & 0.18 & 6725~(60)   & 3600  & 1.1\\
WHT  & 9~Feb~2001  &  2 \times 2048 \times4100~EEV & 0.24 & 6736~(48)   & 6600  & 0.85\\
COMB  &            &                               & 0.38 & ON          & 21600 & 1.9\\
COMB  &            &                               & 0.38 & OFF         & 2700  & 1.9\\
\noalign{\smallskip}
\hline
\end{array}
\]
\end{table*}
\section{$\rm H_{\alpha}$ Observations}

\begin{figure}[!h]
\epsscale{0.8}
\plottwo{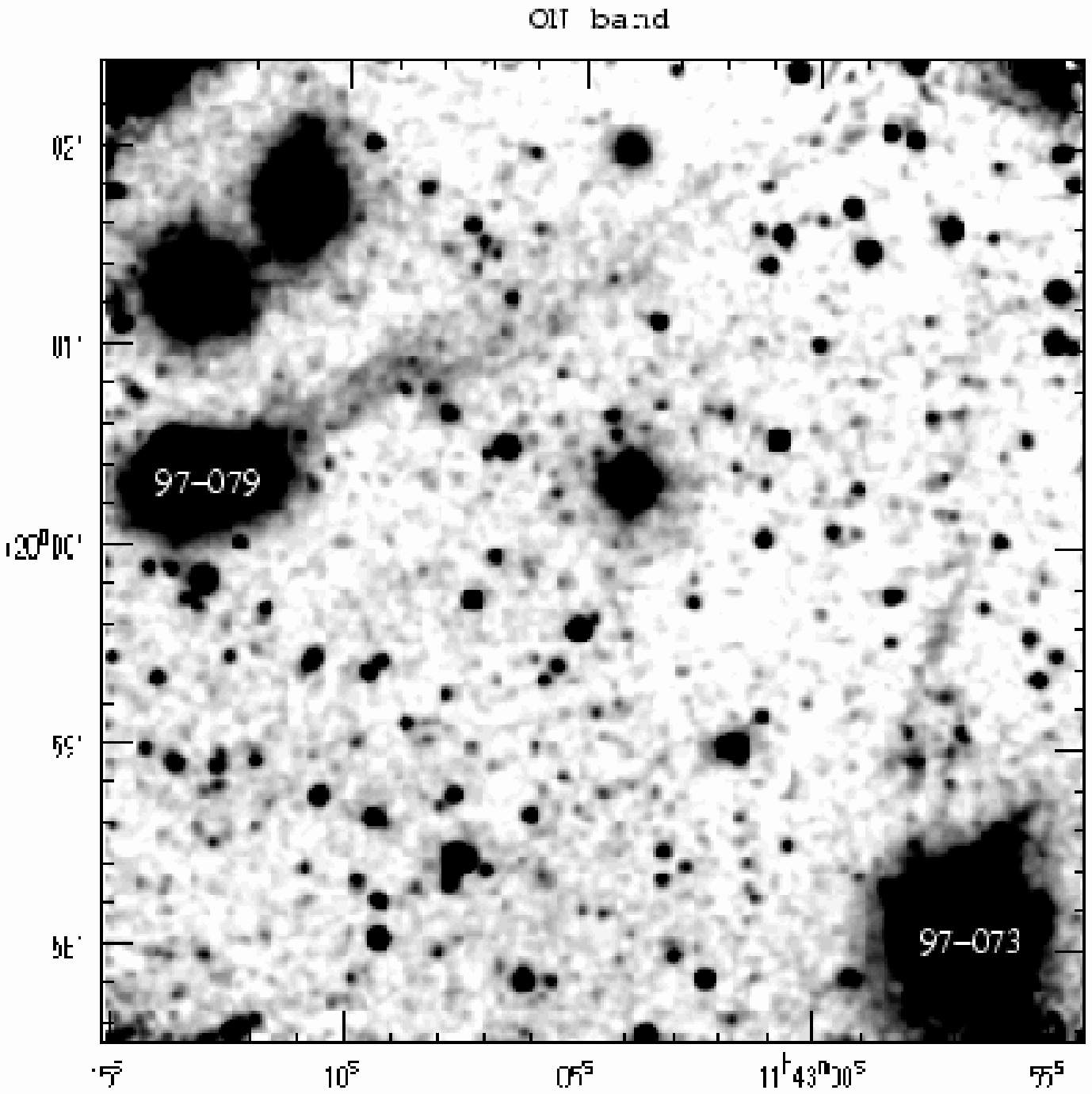}{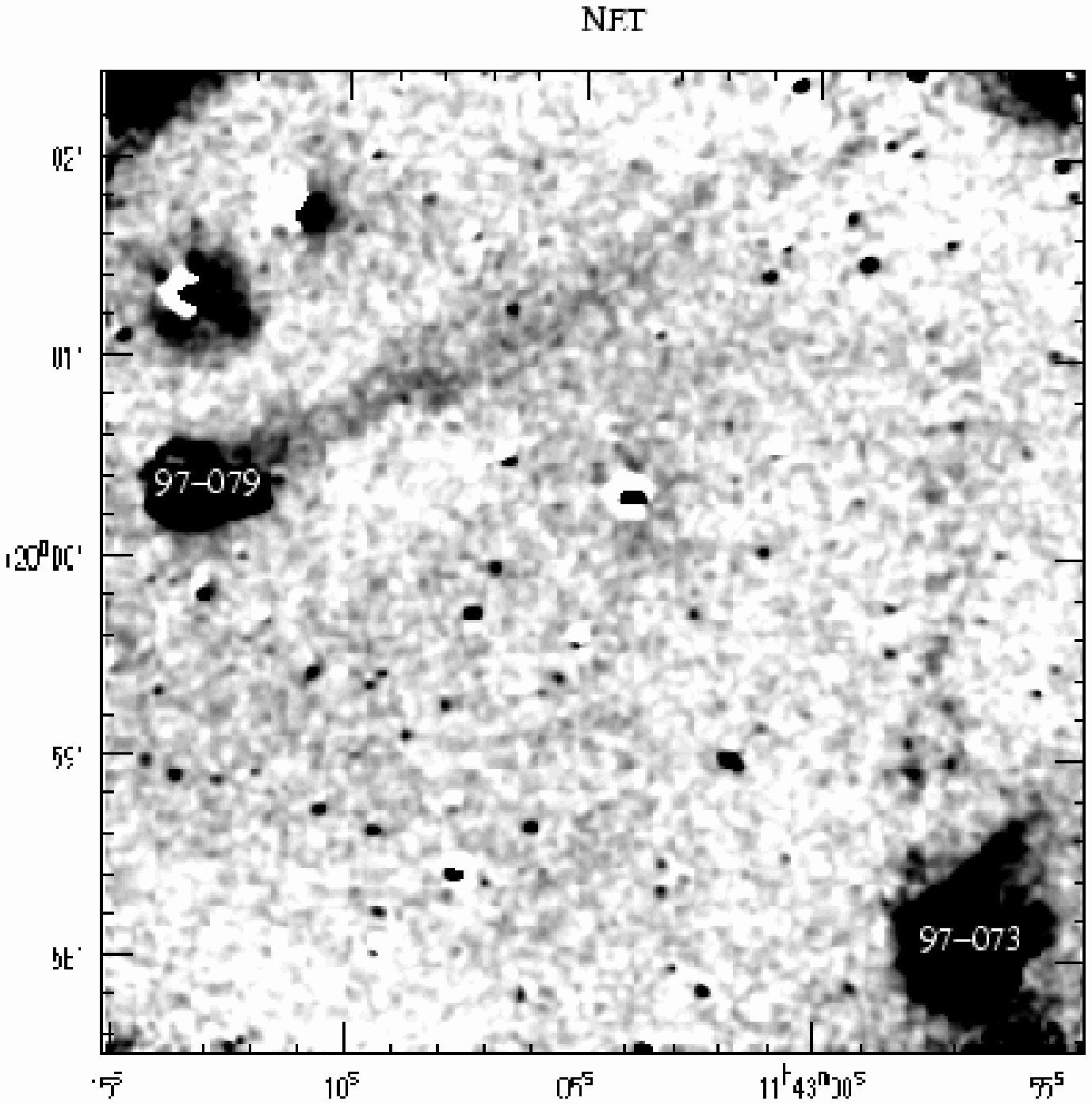}
\small{\caption{The stacked 6 hours $\rm H_{\alpha}+[NII]$ ON band  (left)
and NET exposures (right) of the region containing the two galaxies under study
shown at high contrast to enhance the extended tails.
The upper corners of the images suffer from filter vignetting.}\label{codeON}}
\end{figure}

We observed the field centered on CGCG 97-073 and 97-079 using three
telescopes: the 2.5 m Isaac Newton Telescope (INT), the 2.5 m Northern Optical Telescope (NOT) 
at La Palma (Spain) and the
2.1 m telescope at San Pedro Martir (Mexico) (see Table 1).
The observations were performed through
narrow band filters centered at $\sim$ 6725~\AA,
covering the redshifted $\rm H_{\alpha}$ and [NII] lines.
The underlying
continuum was taken through broad band red filters (see Table 1 for details).
The images were obtained in photometric conditions
with a seeing ranging from 1.1 to 1.9 arcsec.
Each integration was split in shorter exposures (typically 20 min) to get rid of the
cosmic rays. 
The photometric calibration was obtained exposing the spectrophotometric
star Feige 34. 
The individual images were bias subtracted and flat-fielded
using combinations of exposures of several empty fields at twiglight.
After background subtraction, the images were combined.
This was accomplished using IRAF tasks. Based on approximately 100 common stars 
found in the field using DAOFIND, the proper coordinate transformations i.e. rotation, traslation
and re-sampling (independently on X and Y) were applyed using GEOMAP.
The resulting stacked frames, corresponding
to 6 hours and 45 min of integration time (ON and OFF band respectively),
have 0.38$\arcsec$ pixels.
The intensity in the combined OFF-band frames 
was normalized to that of the combined ON-band one using the flux ratio of several field stars. 
The combined NET-image was obtained by subtracting the
normalized OFF-band frame from the ON-band one. The resulting 5$\arcmin \times 5 
\arcmin$ ON- and 
NET-frames are shown in Fig. \ref{codeON}. 
One additional $\rm H_{\alpha}$ exposure of the field was obtained with the 
4.2m William Herschel Telescope (WHT) 
at La Palma. Given the excellent seeing conditions (0.85 arcsec) it provided 
high resolution
images of the two galaxies shown in Fig. \ref{97-073}. 
However, due to the bright moon conditions,
it did not provide any useful information about the low-brightness features, thus it was not
stacked with the remaining observations.

\section{$\rm H_{\alpha}$ tail parameters}
The $\rm H_{\alpha}+[NII]$ fluxes within the optical extent of the two galaxies 
under study are found consistent with 
$\rm Log F(H\alpha+[NII]) = -12.75$  (-12.66) $\rm erg~cm^{-2} sec^{-1}$  
(hereafter the first quantity refers to 97-079, the second one in parenthesis to 97-073)(Gavazzi et al. 1998).
The flux estimate of the low brightness tails requires a carefull assessment of the quality of the  
flat-fielding. For extended sources 
the dominant source of error is associated with the variations of the background on scales similar
to the source.
We measured the background in several $\rm 10 \times 10~arcsec^2$ regions around the field (comparable with the
cross section of the tails) and determined the rms of the background on this scale. 
The flux density in the tails was found to decline from 8 $\sigma$ to 2 $\sigma$
at $\sim$ 3 ($\sim$ 2) arcmin from their parent galaxies. 
Assuming a distance of 86 Mpc to A1367 
the total extent of the tails results in 75 (50) kpc. 
The cross section of the tails is $\sim$ 8 kpc.
The total flux in the trails, obtained integrating the counts above 2 $\sigma$
in rectangular regions of approximately $3 \times 0.3$ arcmin ($2 \times 0.3$ arcmin)
(excluding the galaxies themselves) results in:
$\rm Log F_{trail} = -14.11$ (-14.85) $\rm erg~cm^{-2} sec^{-1}$. 
The line intensity is 0.25 (0.07) Rayleigh
(1 Rayleigh = $\rm 10^6/4\pi~photons~cm^{-2}~s^{-1}~sr^{-1}$), corresponding to an emission measure (EM)
of 0.67 (0.18) $\rm cm^{-6}~pc$ if the gas is optically thin at $10^4$ K.
Assuming that the tails have cylindrical symmetry,  with a filling factor of 1,
this implyes a mass of ionized gas of
$9.6\times 10^8$ ($3.2 \times 10^8$) $\rm M\odot$, thus a plasma density of 
$9 \times 10^{-3}$ ($5 \times 10^{-3}$) $\rm cm^{-3}$ or lower if the filling factor is $<1$.
This estimate is comparable with the present 
content of neutral hydrogen of the two galaxies: $1.6 \times 10^9$ ($2 \times 10^9$) $\rm M\odot$ respectively
(Gavazzi, 1989). 
The plasma in the tails was probably 
ionized inside the galaxies by the actively star forming HII regions on their leading-edges.
However it must have originally consisted of neutral hydrogen, therefore
a rough estimate of the total gas loss from the two galaxies can be derived using the HI deficiency parameter,
as defined by Giovanelli \& Haynes (1985). 
Although one should take this estimate cautiously, because for Irregular galaxies the reference
HI content is highly uncertain, 
from the HI deficiency parameter 0.25 (0.16) of the two galaxies respectively  we
estimate that $1.25 \times 10^9$ ($9.1 \times 10^8$) $\rm M\odot$ of gas was left behind in the tails
or that the two galaxies have lost approximately 40 \% of their original gas. This estimate is
consistent with the mass previously found in the ionized gas.\\ 
Assuming a transit velocity through the cluster of $\rm \sim 1200~km~s^{-1}$, as derived 
from $\sqrt 2 \times \sigma_{vel}$
(here we assume that the cluster can be modeled as an isothermal sphere,
as in Cayatte et al. 1994)
with $\sigma_{vel}=822 ~km~s^{-1}$ for A1367 (Struble \& Rood 1991),
from the length of the tails we 
derive that the ionized material survived some $10^{7.8} (10^{7.6}$) yr.
The recombination time $\tau_r=1/N_e \alpha_A$, where $\alpha_A=4.2 \times 10^{-13}$ $\rm cm^3 s^{-1}$
(Osterbrock, 1989) is $10^{7.0} (10^{7.2}$) yr i.e. about 5 times shorter than the survival time.
The two times would however become consistent if the plasma density was about 5 times lower 
than the value estimated above, depending on the filling factor.
Although it is likely that the gas in the tails came out ionized from the galaxies,
it cannot be excluded that 
the eddies along the tails contain sufficient turbulent energy to currently sustain
its ionization along the tail.

\section{Discussion}

Both 97-073 and 97-079 appear to have 
lost to the tails  $\sim 40 \%$ of their original gas content.
Stripping of the gas component of galaxies in clusters can occur either 
due to the ram pressure exerted by the intracluster medium
(Gunn \& Gott 1972) or because of tidal interactions. 
Ram pressure acts on a very short timescale, of the order of $10^7$ yr, 
and produces a stream of gas that trails the galaxy 
(Abadi et al. 1999; Quilis et al. 2000).
Tidal interactions within galaxy clusters are of two different types;
(a) galaxies interact with the global potential of the cluster, being tidally shocked each time they
approach the pericenter of their orbit inside the cluster core (Merritt, 1983), and (b) they interact with the other
galaxies through repeated high-speed fly-by encounters, a mechanism known as galaxy harassment (Moore et
al. 1996, 1998, 1999). Global tidal interactions can
remove a large fraction of the gas reservoir of galaxies
(Mayer et al. 2001a, b) but they require a long timescale,
of the order of 7-10 Gyr, which is longer than the typical age of galaxy clusters (Rosati et al. 2000).
Galaxy-harassment acts on a much shorter timescale
in clusters; the collision rate per galaxy in
a rich cluster like Coma is $\rm \sim 1/Gyr$ and 1-2 
collisions with a large $L_*$ galaxy might remove $\sim
50\%$ of the gas in a disk galaxy (Moore et al. 1998). 
However, harassment, as any other tidal
mechanism, is expected to produce a leading gaseous tail in addition to the trailing tail, 
which can be excluded for the galaxies here analyzed. 
Overall, ram pressure seems to provide the most likely explanation for the features observed.

\smallskip
Yet, given the density of
the gas in the NW subcluster of A1367, namely
$\rho_g \sim 7 \times 10^{-4}$ atoms~cm$^{-3}$
(Donnelly et al. 1998), a factor $\sim 5$ lower than the density in the 
core of Coma, ram pressure is expected to be rather weak according to recent
numerical simulations (Abadi et al. 1999; Quilis et al. 2000). However,
these studies focused on the effects on large spiral galaxies 
whose potentials are much more centrally 
concentrated and should yield stronger restoring forces
than those of the faint, irregular galaxies considered here.
Moreover, the relative positions and present apparent direction of motion of the galaxies (we
assume that the galaxies are moving in the opposite direction with respect to the gaseous tails)
are consistent with them having undergone a very close encounter; 
assuming that their
velocity is  $\rm \sim 1200~km~s^{-1}$, the encounter likely took place some $10^{7.8}$ yr ago.
The tidal encounter could have lowered the restoring force by loosening
the potential well of the galaxies; we tested this hypothesis with
a few N-Body simulations performed with the 
binary treecode PKDGRAV (Dikaiakos \& Stadel 1996; Stadel \& Quinn,
in preparation). The target galaxy is represented by a high-resolution
N-body model analogous to those described in Mayer et al. (2001a, b), that obey
the Tully-Fisher relation (Zwaan et al. 1995), and 
comprises an exponential stellar disk of 50,000 particles embedded in 
an isothermal dark matter halo of 400,000 particles.
The model has structural parameters and a luminosity 
consistent with those measured in 97-073 and 97-079.
The perturbing galaxy is represented by the external potential of
an isothermal halo with mass comparable to that of the target model.
The relative velocity of the galaxies is varied in the
range  $V_{rel}= 500- 1000$ ~km~s$^{-1}$ (i.e. comparable to the velocity
dispersion measured for the galaxies in the cluster, see section 3), the impact 
parameter is equal to the core radius of the perturber and
various disk orientations
are considered. We analyze the target galaxy $\sim 10^8$ yr
after the collision in the various runs and
find that its mass distribution, and thus its restoring
force, has been only slightly affected, both because
the perturbations are intrinsically small and because the response
of the system is slow
compared to the time elapsed since the
collision has occured. However, if the galaxies travel at $\sim 1200$
km/s in the cluster, the ram pressure would overcome their restoring
force at more than the two disk-scale lengths
from the center, mainly because their potential is intrisically
shallow. Stripping down to $2$ scale lengths would remove $\sim 25 \%$ of the
gas if it were distributed as the stellar disk; turbulence 
and viscosity would enhance gas stripping (Quilis et al. 2000) and, 
in addition, gaseous disks in late-type
galaxies are usually more extended than the stellar disk of our model (de Blok
\& McGaugh 1997) and would be more fragile.
Hence, unless 97-073 and 97-079 have potentials 
much more concentrated than assumed here, our results are consistent
with the hypothesis that ram pressure has produced their gaseous tails.

\acknowledgements

G.G. wishes to thank Christian Bonfanti for his contribution to the image 
reduction and Luca Cortese for useful discussions. L.M. thanks Thomas Quinn  
and Joachim Stadel for allowing him to use PKDGRAV and Ben Moore 
for useful discussions. L. Carrasco research is supported by
CONACYT research grant G28586-E.

\smallskip

\end{document}